# USER PROFILE BASED PROPORTIONAL SHARE SCHEDULING AND MAC PROTOCOL FOR MANETS

Hannah Monisha . J[1] and Rhymend Uthariaraj .V[2]

[1]Indira Gandhi College of Arts and Science, Govt. of Puducherry, India.
monisha_cyril@yahoo.com
[2]Anna University, Chennai, India.
rhymend@annauniv.edu

## ABSTRACT

*Quality of Service(QoS) in Mobile Ad Hoc Networks (MANETs) though a challenge, becomes a necessity because of its applications in critical scenarios. Providing QoS for users belonging to various profiles and playing different roles, becomes the need of the hour. In this paper, we propose proportional share scheduling and MAC protocol (PS2-MAC) model. It classifies users based on their profile as High Profiled users (HP), Medium Profiled users (MP) and Low profiled users (LP) and assigns proportional weights. Service Differentiation for these three service classes is achieved through, rationed dequeuing algorithm, variable inter frame space, proportionate prioritized backoff timers and enhanced RTS/CTS control packets. Differentiated services is simulated in ns2 and results show that 9.5% control overhead is reduced in our proposed scheme than the existing scheme and results also justify that, differentiated services have been achieved for the different profiles of users with proportionate shares and thereby reducing starvation.*

## KEY WORDS

*Proportional-share, QoS, MANET, Differentiated services, MAC protocol*

## 1. INTRODUCTION

Recently, usage of mobile devices such as third generation mobile phones, Personal Digital Assistants (PDAs) and laptops are in the rise. They are becoming increasingly popular because they support many applications from simple message transfer to complicated location based services[1]. The infrastructure-less Mobile Ad Hoc Networks will be the most widely used network of the future[2].

A MANET is dynamically formed when more than one individual mobile device want to interact for temporary short duration with others for any purpose such as accessing application or data transfer. Every node in a MANET is an independent light weight device which can move randomly. This makes their transmission range limited and network topology dynamic[3]. All the nodes in the MANET take the role of host for sending and receiving and intermediate node for routing purposes. Since their nodes are autonomous, there is no central control. The nodes self organize and collaborate among themselves. Further, the user of the mobile nodes are not restricted to be a part of one MANET alone but can be connected to other MANETs or public networks such as Internet. The above mentioned features of MANETs, makes it more complex to configure. Since the member nodes of MANETs are battery operated, they impose power constraints. Further reliability is reduced by error-prone, unstable and asymmetric nature of the wireless links. MANETs find its application in business and commercial applications, electronic class rooms, convention centres, emergency rescue operations (fire, flood, earthquake etc), law enforcement (crowd control) and military applications. Specialized MANETs are Personal Area

Networks, Residential Mesh Networks, Vehicular Adhoc Networks, Wireless Sensor Applications etc [4]. Since MANETs find its application in critical scenarios providing QoS becomes necessary. Thus there is need to establish new architectures and services. Certain applications require differentiated services architecture that offers multiple service levels, each with different QoS requirements. Implementing QoS to wireless nodes is complex due to its mobility and ad hoc nature.

The vital resource in a wireless network is the bandwidth. Hence allocation of this bandwidth to various nodes has to be done judiciously. Efficient scheduling algorithms to resolve contention between various contending nodes and efficient bandwidth utilization in differentiated services environment derive prime attentions. Throughput and fairness are the two QoS parameters to be considered in differentiated services. Fairness is an important issue when accessing a shared wireless channel. With fair scheduling, different categories of users demanding various levels of QoS wishing to share the wireless medium can be allocated proportionally.

The ability to provide QoS is mainly dependant on how well the resources are managed at the Medium access Control (MAC) layer. The contention for channel access is resolved at the MAC layer. The Carrier Sense Multiple Access Protocol with Collision Detection(CSMA/CD) insists on carrier detection and a node is required to sense the wireless medium for a specified time duration before transmission. The CSMA/CD also employs collision detection mechanism to enhance medium utilization. Although CSMA/CD has been proven successful in wired networks, it cannot be directly ported to wireless networks due to hidden terminal and exposed terminal problems. Enhancing MAC protocols to handle these issues and improve the performance of MANET applications and provide efficient QoS is a challenge.

Generally a network supports various types of data flows which include real-time traffic such as voice and video and non-real-time traffic such as messages. In a differentiated services environment, the real-time traffic is normally expected to get better service than the others. Hence priority for acquiring resources in a network is generally given to the real-time traffic. In MANETs, in addition to real-time flows, non-real time flows such as messages play a very crucial role because of their application in emergency rescue operations and military battle field[4]. These applications require fast and assured transmission of emergency messages. Further the profile of the user sending the message is also important because they follow hierarchical organizational structure[5]. Hence in such a scenario, priority for acquiring resources in a network has to be given to the emergency non-real-time traffic based on the user profile.

The main objective of this research paper is to define a MAC model that supports QoS based bandwidth utilization based on proportional sharing. This would avoid starvation among the low priority nodes ensuring them a proportional share of bandwidth. This paper proposes a proportional share scheduling and MAC protocol (PS2-MAC) model based on user profile for MANETs. The model incorporates, rationed dequeuing algorithm to favor proportional share, variable inter frame space, proportionate prioritized backoff timers to resolve channel contention and enhanced RTS/CTS packets to support priority. Variable retry counters are used to avoid starvation. Priority reversal is also avoided with the help of enhanced CTS. The paper is organized as follows. Section 2 presents review of literature; Section 3 explains the proposed model, followed by simulation results in Section 4 and Section 5 gives conclusion and future directions.

## 2. REVIEW OF LITERATURE

QoS is a set of service requirements to be met by the network[3]. QoS in internet has been extensively studied by researchers. Since MANET may also require internet connectivity to use certain applications, QoS solution for internet can be considered for MANETs[3]. Integrated services (IntServ) approach[6] proposes three service classes namely guaranteed service, which offers high QoS, controlled load service, which offers medium QoS and Best Effort service, which offers low QoS. IntServ model proposed hard QoS and it is difficult to achieve in MANET because of MANET's dynamic topology and limited resources[2]. The other model is Differentiated Services(Diffserv)[7]. Diffserv classifies the traffic into three services. They are Premium service, Assured service and Best effort service. Diffserv is not very suitable for MANETs because as it does not guarantee services on per hop basis[3]. AQOR[8] is a QoS model which was proposed based on reservation. FQMM[9] is a QoS model which uses IntServ for high priorities and Diffserv for Low priorities. Still problems such as, decision on traffic classification and scheduling has to be addressed[3]. Few QoS models for MANETs are surveyed in [10]. Each of these algorithms have their own merits and demerits.

Some QoS models in MANET deal with adapting IEEE 802.11e standard[11]. It supports differentiation at the MAC layer using EDCA for contention and HCCA for polling. EDCA can be considered for MANETs but the high priority traffic often suffers delay due to alternate blocking problem[12]. Moreover, 802.11e follows a priority queue model where, only when the high priority queue is empty, the next higher queue is dequeued. This creates starvation among the low priority queues. Another drawback is that, classification is based on the type of the data being transmitted and no preference is given for the user or the urgency of the data [13]. [14] explains clustering based on user profile and build trust. [15] Discusses various weight based algorithms to support selection of leader in a cluster of nodes. Weights are assigned based on various network parameters. [16] proposes a context aware algorithm where, when a node faces degradation in throughput, the packets are handed off to a less loaded node to improve QoS. [17] proposes a scheduling algorithm similar to 802.11e where the buffer space allotted to the low priority is restricted, which leads to starvation of low priority traffic and unfairness. Apart from scheduling a suitable admission control mechanism also influences QoS[18]. Users seeking High QoS are allocated bandwidth first. The users who are assigned low QoS may be blocked to an accepted level if all the bandwidth has been used by existing High QoS users. But this may lead to starvation among low QoS users. [19] proposes differentiation based on channel conditions. Though in recent times number of protocols has been proposed to support QoS, fairness has not been considered [20].

The priority access service (PAS) was developed by the Federal Communications Commission(FCC) and is managed by the National Communication System(NCS). The primary mission of PAS is to support National Security and Emergency and Preparedness (NS/EP) telecommunication services by providing services such as Wireless Priority Service (WPS) in a mobile environment, which will allow military and civilian personnel to access cell channels ahead of general public in times of crisis. Further PAS provides different levels of priority with each level having a distinct class of users categorized based on their roles[5]. These initiatives, further strengthens our motives for the need of prioritization in MANETs based on user profiles.

Classification of users, based on user profiles, is also studied from other related services. In [21] WiMAX users are classified into three categories Platinum, Gold and Silver. In [22] Cloud users are categorized as Class A(High QoS), Class B(Medium QoS) and Class C(low QoS). In [23] Grid customers are segmented as Premium, Business and Budget users. This very well shows that three categories of QoS are required to attain customer satisfaction.

D-MACAW[24] proposes differentiated services based on user profile. It categorizes the users as High privileged and low privileged. Drawback in D-MACAW is that they simply assign twice the bandwidth allotted to low priority to the High priority. They follow only node priority. They do not consider the packet priority. When the packet belonging to the high priority node is forwarded by a low priority node, it is treated as low priority packet at that time. This leads to priority reversal of the packet. Other demerit is that, users who wish to choose between high and low QoS have no intermediate option. Having only two QoS classification limits the user's choice. The same authors proposed AT-ST scheme[25], where they add two more control packets Alert Transmission(AT) and Suspend Transmission(ST) to enhance differentiation.

Further in AT-ST scheme, adding two more control packets in addition to the existing control packets would create control packet overhead. This would result in degradation of throughput in a MANET.

Hence, we enhance the D-MACAW. We propose PS2-MAC model based on user profile for MANETs. We classify the users as High profiled user with good QoS, Medium profiled user with moderate QoS and Low profiled user with Best effort service. Further we assign individual weights to the users to differentiate them proportionally. We incorporate rationed dequeuing algorithm to resolve starvation in prioritized scheduling, differentiated waiting times to enhance priority and an alternate solution to support three priority and to overcome the control packet overhead occurred with AT-ST scheme by . This takes care of the priority reversal problem also.

## 3. PROPOSED MODEL

### 3.1. Classification of users

To meet the user's service requirements, we propose to classify the users as High Profiled user(HP) with good QoS, Medium Profiled user(HP) with moderate QoS and Low Profiled user(LP) with Best effort service. Every user is assigned a static priority according to the classification HP, MP and LP based on their user profile or requested QoS level. Every user is assigned a proportional weight to meet their service requirements. This proportional weight favors fairness among the competing nodes in a differentiated services environment. To implement this, we add an additional field called Priority Field to the header of every packet that is generated at the source node to store the priority of the node. The codes 0, 1 and 2 representing the priorities HP, MP and LP are stored in the priority field. A node acts as both source and intermediate node. The source node does packet stamping where the priority of the user is stamped in the priority field of the packet and the intermediate node just does the enqueuing based on priority field.

### 3.2. Rationed Dequeuing Algorithm

Scheduling has always been a way to resolve channel contention. It is the technique that decides who acquires the channel next. Our objective is to allocate proportional share of bandwidth among the contending nodes. To achieve this, at every node we maintain separate queues HP, MP and LP for the three classes of users. The packets are enqueued in their respective queues according to their priority mentioned in the Priority Field. The packets are dequeued from the queues proportionally based on their weights and the percentage of packets waiting in their respective queues. Here the weights are constant and the percentage of queue length(QL) is dynamic. The percentage of Queue length x, y and z for HP, MP and LP queues respectively are calculated as in equation (1).

$$x = \frac{QL_{HP}}{QL_{HP} + QL_{MP} + QL_{LP}} \text{ ,}$$

$$y = \frac{QL_{MP}}{QL_{HP} + QL_{MP} + QL_{LP}} \text{ ,}$$

$$z = \frac{QL_{LP}}{QL_{HP} + QL_{MP} + QL_{LP}} \qquad (1)$$

We decide the number of packets to be dequeued, from each queue based on their access ratio given in equation (2).

$$w_0 x : w_1 y : w_2 z \qquad (2)$$

Where, $w_0$, $w_1$, $w_2$ are the user defined weights assigned for HP, MP and LP nodes such that $w_{max} > w_0 > w_1 > w_2 > 0$, $w_{max}$ is the maximum weight that can be assigned. Weights play a vital role in deciding the proportion in which the user profiles are differentiated. Hence care should be taken while assigning the weights to ensuring fairness among the different user profiles. x is the percentage of HP packets, y is the percentage of MP packets and z is the percentage of LP packets waiting in their respective queues. When the traffic is dominated by MP or LP packets, the queue length of MP and LP may increase and thus the access ratio of HP packets may fall below MP or LP. Hence to ensure the priority of the HP node, when the percentage of HP packets fall below the average percentage, we maintain the percentage of HP packets at an average Av=100/3≈33. Similarly if the percentage of LP packets becomes greater than the MP packets, then their percentage is maintained at the average percentage. Thus at any point of time, for any random data flow, the priority of HP packets is ensured. Similarly, the priority of MP over LP is also ensured. The following algorithm(1) is designed to achieve this.

| Algorithm I : Rationed Dequeuing Algorithm |
| --- |
| Step 1 : If (x<Av) then x = Av |
| Step 2 : If ((y>x) or (y<z)) then y= Av |
| Step 3 : If ((z>y) or (z>x)) then z= Av |
| Step 4 : Access ratio= $w_0.x : w_1.y : w_2.z$ |

Calculating access ratio based on weights and Queue length, introduces a level of dynamicity in the proportional share scheduling. Further fairness is ensured with avoidance of priority reversal that may occur due to increase in percentage of LP over MP or HP.

### 3.3. Proportional Prioritization at the IEEE 802.11 MAC Layer

Once the packet is dequeued and ready for transmission, the next step is to acquire channel access. IEEE 802.11 for wireless LANs is the widely used MAC protocol. The 802.11 distributed coordinated function(DCF)[26] serves both infrastructure and ad hoc architectures. Every contending station has to go through a contention resolution procedure to determine which station can transmit next. Once a node wins the contention, it waits for a backoff time and sends a request to send(RTS) message to the intended receiver. On reception of the RTS, the receiver replies with a clear to send(CTS) message. On reception of CTS, source forwards the data packets. On reception of the data packets, the receiver sends an acknowledgement (ACK). When the current transmission is successful, the contending station waits for an inter frame space and then a new round contention for the medium begins. There are two waiting stages during contention, the Inter Frame Space(IFS) and the Back-off stage. The priority differentiation at the channel contention is achieved at the IFS by equation (3) adapted from the IEEE 80211e.

$$IFS_i = SIFS + AIFSN_i * slot\ time;\ i=0\ to\ 2 \qquad (3)$$

Where, SIFS is the Short Inter frame space, AIFSN is the Arbitrary Inter Frame Space Number. We propose to differentiate the AIFSN proportionally based on the weights. It is calculated using formula (4). The higher the weight assigned to a node, the lower will be the AIFSN. Hence, the IFS will be shorter and thus the priority will be higher.

$$AIFSN_i= integer\left(\frac{\sum_{j=0}^{2} w_j}{w_i}\right)\ ;\ i=0\ to\ 2 \qquad (4)$$

The next stage is the Back-off stage. 802.11 DCF uses a backoff counter at each node such that every node can choose a random number between zero to maximum contention window size. After sensing the channel to be idle for an IFS, the nodes start counting their backoff counters to zero. If the channel is found to be busy, they freeze their backoff counters. The value of contention window is assigned between minimum and maximum contention window. IEEE 802.11 DCF is a random access mechanism, where a node selects a backoff value based on the formula (5).

$$Backoff = integer(2^{\,2+k} * random() * slot\text{-}time) \qquad (5)$$

Where random() is the random number evenly distributed between 0 and CW, where CW is the Contention Window which varies between minimum($CW_{min}$) and maximum contention window ($CW_{max}$) and $k$ is the number of attempts made for transmission.

To further support prioritization and to reduce collision, differentiated backoff timers for HP, MP and LP are proposed, proportionate to their access ratios, using formula (5) as in the following equation (6) adapted from [27].

$$Backoff\_Prio = integer(PF_i^{\,2+k} * random() * slot\text{-}time) \qquad (6)$$

Where, $PF_i$ is the priority factor for HP, MP, LP. The authors propose user defined priority factor. Since we already have user defined weights, we calculate PF proportionate to the weights[28]. The higher the weight assigned, greater will be the share of bandwidth allocated. The lower the PF, lower will be the waiting time. Hence, PF should be such that $0 < PF_0 < PF_1 < PF_2 < 1$. The following formula (7) calculates the PF for HP, MP, and LP proportional to their weights.

$$PF_i = 1 - \frac{w_i}{\sum_{j=0}^{2} w_j} \quad ; i{=}0 \text{ to } 2 \qquad (7)$$

Once, the backoff reaches zero, before the data is transmitted, the source station sends a RTS and receives a CTS following which it transmits DATA and gets an ACK. RTS-CTS-DATA-ACK transmissions takes place. RTS, CTS and ACK are called control packets. In the event of not receiving a CTS or ACK, the source is led to believe that a collision has occurred. In order to avoid further collisions, the backoff timer is increased. The $CW_{min}$ doubles after every collision till it reaches $CW_{max}$. The RTS-CTS exchange is optional in 802.11. But, it is widely used with wireless networks to avoid the hidden terminal problem.

Figure 1 explains the proposed model PS2-MAC, which shows the working at one node which can act as a source or intermediate node. Once a packet is generated, packet stamping is done according to the user profile based priority in its priority field. If it is a forwarded packet no packet stamping is done. The packets are enqueued in the three different queues HP, MP and LP based on the stamping in the priority field. Then, according to rationed dequeuing algorithm, access ratios are normalized and the packets are ready for dequeuing. Based on proportional share scheduling algorithm, contention among the three queues are resolved and finally a packet is ready for transmission. This packet contends for channel access with the packets of other nodes.

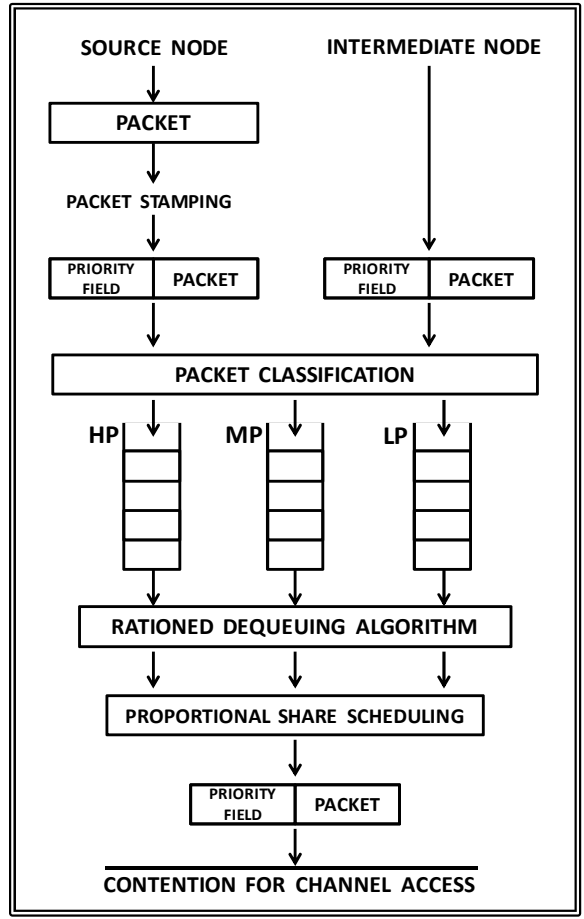

Figure.1. PS2-MAC Model

To resolve contention between nodes and achieve prioritization, [25] proposes AT-ST scheme. This scheme uses two additional control packets AT and ST. This scheme supports only two priorities. The high priority nodes send AT packet to inform the neighboring nodes about the high priority packet transmission. The node that receives the AT packet, checks the backoff value of the sender and compares with its own. If the receiver is of high priority, it will immediately send a ST packet to suspend transmission to the sender node. This above mentioned scheme does not support three priorities, since priority is assigned only through backoffs and variable backoffs are designed only for two. Hence we modify this scheme to support three priorities. Further our scheme also avoids the extra control packet overhead caused by the AT and ST packets.

We integrate the packet priority along with the RTS packet. To achieve this, we add an additional priority field to the RTS packet and store the packet priority analogous to [29]. The priority field values 0,1 and 2 are used to represent HP, MP and LP respectively. Similarly, an additional flag field is added to every Clear To Send (CTS) packet[29]. The flag values 0 and 1 are used to represent 'clear to send' and 'suspend transmission' respectively. Priority reversal occurs when a low priority node has its backoff at zero when the high priority node is in contention. This can lead to a situation where the low priority node acquires the channel before a high priority node. This is resolved using suspend transmission.

When the backoff of a node reaches zero, RTS packet is transmitted. The node receiving the RTS packet follows three steps:

Step 1: it forwards the RTS packet to its neighbors.

Step 2: Checks for the priority of the packet of the sending node in the priority field.

Step 3: If the priority of the sending node is greater than or equal to its own packet priority, it sends a CTS with flag value 0, thus informing to proceed with data transmission. If otherwise, it sends a CTS with flag value 1, thus informing that a priority reversal has occurred, hence defer transmission.

The node receiving CTS follows two steps:

Step 1: if a node receives CTS-0, it continues transmission with DATA and if a node receives CTS-1 defers transmission.

Step 2: If a neighboring node overhears CTS-0 it defers transmission. If it overhears CTS-1 it resumes its state.

This overcomes the problem of priority reversal among the contending nodes and hidden terminal problem to a great extent. Figure (2) explains this. Further, to avoid starvation among low priority nodes during channel contention, [25] proposes retry counter. To support three priorities, we enhance [25] by introducing variable retry thresholds for MP and LP nodes. After every unsuccessful contention, the retry counters are incremented. Once the retry counters reach a retry threshold, priority is given to that node for transmission. Thus the priority of the HP packet is ensured at all levels and starvation of MP and LP nodes is avoided. This method can also be extended to support any number of priority classes.

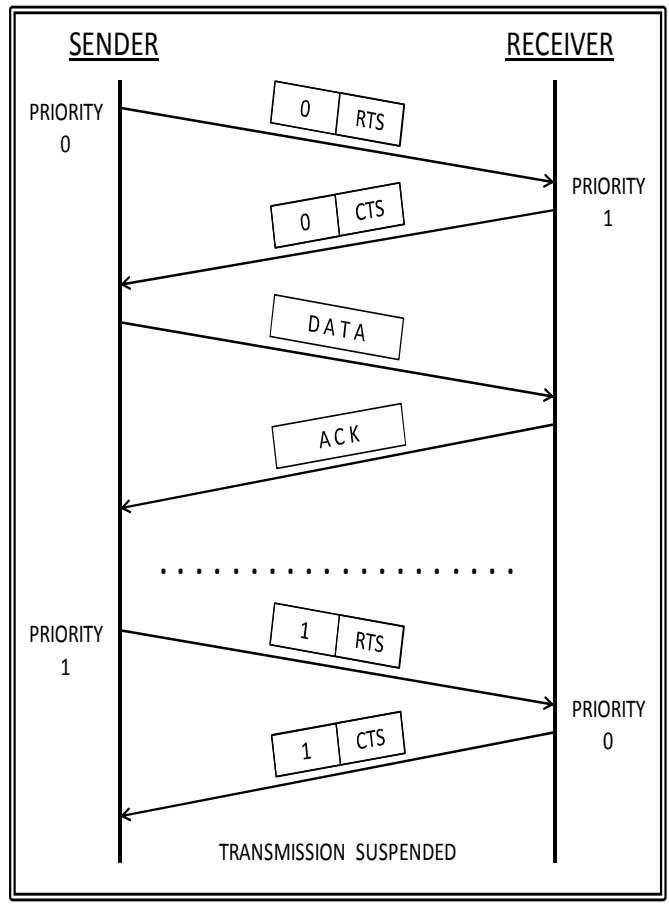

Figure.2. Enhanced RTS – CTS in PS2-MAC

## 4. SIMULATION AND RESULTS

Similar to wired networks, QoS in MANET can be measured in terms of throughput, delay, packet loss, jitter, packet delivery ratio etc. We implemented PS2-MAC in ns2. The test network included 36 nodes each assigned priorities such as HP, MP and LP randomly. The transmission range of each node is defined as 250m and the bandwidth of the channel is 2 Mbps. The DSR protocol is used for routing. For the purpose of simulation, we have assigned the weights for HP, MP, LP as $w_0= 3$, $w_1=2$, $w_2=1$ [30]. Five different scenarios were simulated altering the traffic conditions to study the performance of the model.

Scenario I-    HP traffic dominates the network with very less MP and LP traffic. Such that x>y=z.

Scenario II-   HP traffic is marginally greater than MP traffic and MP traffic is marginally greater than LP traffic. Such that x>y>z.

Scenario III- The medium is shared by equal number of HP, MP and LP traffic. The percentages of packets waiting in the queues are equal. Such that x=y=z.

Scenario IV- MP traffic is marginally greater than HP traffic and LP traffic. Such that x<y>z.

Scenario V- LP traffic dominates the network with very less HP and MP traffic. Such that x=y>z.

We apply algorithm (1) so that the access ratio is always maintained such that access ratio of HP>MP>LP as in table (1). Other parameters are applied as in table (2). The QoS parameters for every scenario were recorded and analyzed. Multiple simulations were run for the same scenario and results were averaged to improve consistency.

Table 1: Access ratio

| Scenario | %HP  x | %MP y | %LP z | Access ratio |
|----------|--------|-------|-------|--------------|
| 1 | 80 | 10 | 10 | 24:2:1 |
| 2 | 50 | 30 | 20 | 15:6:2 |
| 3 | 33 | 33 | 33 | 10:7:3 |
| 4 | 30 | 50 | 20 | 10:7:2 |
| 5 | 10 | 10 | 80 | 10:7:3 |

Table 2: Other Simulation Parameters

| Parameters | Values |
|------------|--------|
| $w_0$ | 3 |
| $w_1$ | 2 |
| $w_2$ | 1 |
| $CW_{min}$ | 32 |
| $CW_{max}$ | 1024 |
| Slot-time | 20 µS |
| SIFS | 10µS |

**4.1.Throughput**

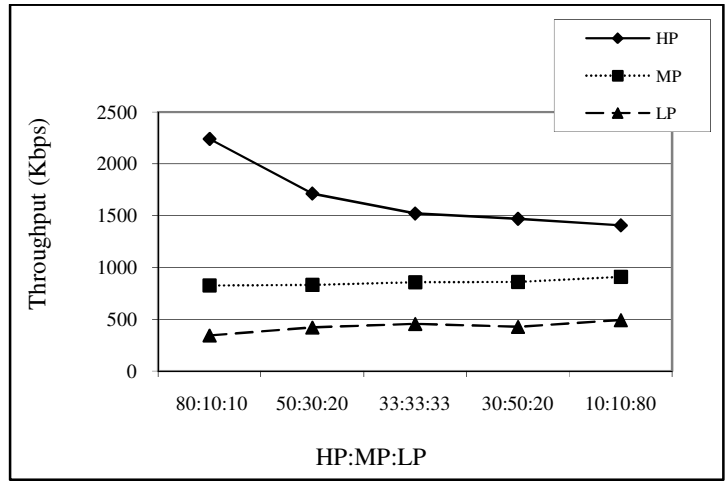

Figure.3. Throughput for HP, MP and LP users

Throughput is calculated as the total number of bits received at the destination divided by the total transmission time. We observed the throughput for the five scenarios. We ran the simulation ten times and aggregated the results. Figure 3 shows that the throughput that was observed during the simulations. It shows that during Scenario I, the throughput of HP increases because of the increase in percentage of HP packets. With regards to MP and LP, the throughput of MP is greater than LP even if the percentage of packets equal. This is because of the proportional weight of MP which is greater than LP. Similarly for Scenario II and III, the throughput of HP>MP>LP. During Scenario IV, the throughput of HP packets does not drastically decrease lower than MP and LP, when the percentage of HP packets is less than the MP and HP packets. This is because of our algorithm (1), where we maintain the access ratio of the HP packets even when it drops below Av. Similarly the throughput of MP and LP does not drop very low, even if their percentages are less because of their fair share allotted through their weights. This also avoids extensive starvation of MP and LP packets. During Scenario V, even if the percentage of HP and MP packets are very low, an average share of the bandwidth is allocated. The throughput of LP has increased because of the increase in the number of packets.

**4.2. Packet delivery ratio**

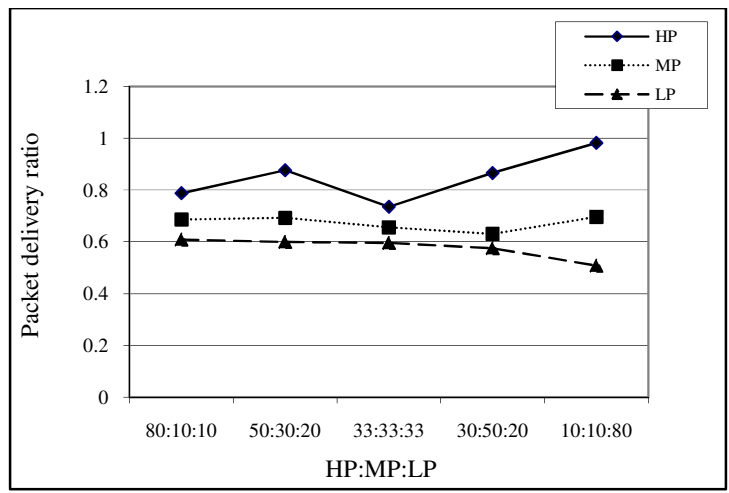

Figure.4. Packet Delivery Ratio for HP, MP and LP users

Packet delivery ratio is calculated as the ratio of the data packets delivered to the destinations to those generated by the CBR sources. Results of five simulation runs were aggregated. Figure 4 depicts the comparative packet delivery ratio. During Scenario I, the packet delivery ratio of HP decreases because of the increase in the percentage of HP packets. This is because, when the percentage of HP packet increases, there is a competition among them, hence packet collision and packet dropping occurs leading to decrease in packet delivery ratio. In Scenario II, the percentage of HP decreases hence the packet delivery ratio has improved. In Scenario III, percentage of nodes of all three priorities are equal. Hence the packet delivery ratio is proportional according to their weights because of proportional sharing. In Scenarios III and IV, when the percentage of HP packets is less, channel utilization is possible without much competition among them; hence higher packet delivery ratio is achieved. When the percentage of HP packet increases, there is a competition among them, hence packet collision and packet dropping occurs leading to decrease in packet delivery ratio. Similarly the Packet delivery ratio of MP and LP nodes decreases with increase in the percentage of nodes. At any point of time during the simulation, it is observed that the packet delivery ratio of HP>MP>LP. This is because of the rationed dequeuing algorithm. Similarly the packet delivery ratio of MP and LP does not drop very low, even if their percentages are less because of their fair share allotted through their weights and the retry counters which ensures transmitting packet before it is dropped. This also avoids extensive packet drop of MP and LP nodes which is generally experienced with other priority scheduling algorithms. In Scenario V, even if the percentage of LP is very high resource is rationed using rationed dequeuing algorithm and priority reversal is avoided using enhanced RTS-CTS.

### 4.3. Average end-to-end delay of data packets

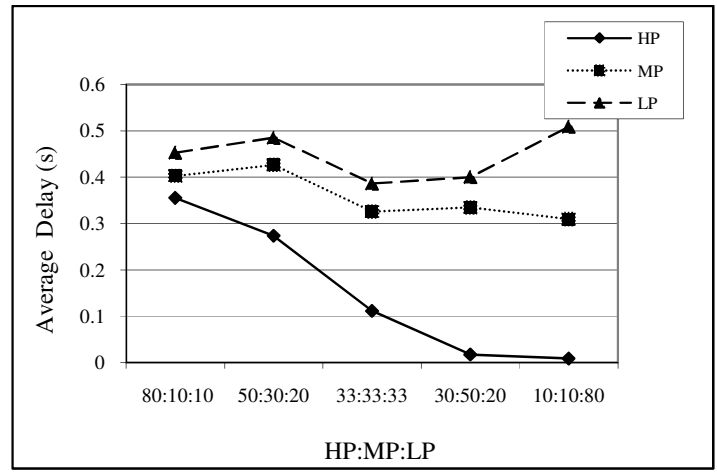

Figure.5. End-to-End Delay for HP, MP and LP users

End to end delay includes, all possible delays caused by buffering during route discovery latency, queuing at the interface queue, retransmission delays at the MAC, and propagation and transfer times. Average delay is calculated as, the average of the difference between the time when the packet is received by the destination, and the time it has been sent from the source. Aggregated simulation results are illustrated in Figure 5. In Scenario I, HP faces the maximum delay. The average delay increases with the increase in percentage of HP packets and decreases with the decrease in percentage of HP packets. Even when the percentage of HP is more, the delay experienced by the HP packets is less than the MP and LP. This is because we have prioritized at all levels such as backoff, IFS and at contention. Thus the overall performance of HP is superior followed by MP and finally LP. In Scenario II, III and III when the percentages of the packets are almost the same, the delay of HP is very less. In Scenario III, when the percentages of packets are equal, the delay is least for LP. This is because of the proportional share allotted to introduce fairness and avoid starvation. In Scenario V, The delay of HP is almost zero when it is very less and the delay of LP increases when it dominates the network.

### 4.4. Control Packet Overhead

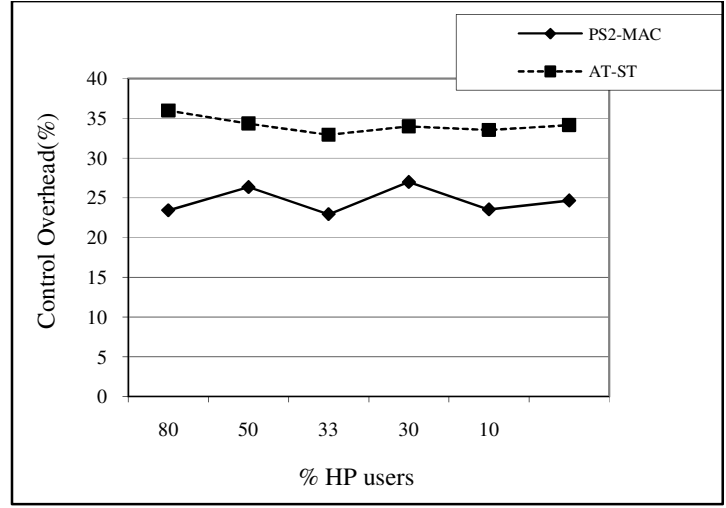

Figure 6: Comparison of Control overhead

We compare our model with the existing model AT-ST. To prove that the performance of our model is better than the existing model, we simulated AT-ST in ns2 and observed the performance. Since AT-ST supports only two priorities, comparison was done based on control overhead. Instead of the additional control packets AT and ST that was used additionally with RTS/CTS, we propose enhanced RTS-CTS where we embed control flags along with RTS-CTS control packets. The control packet overhead is calculated as the total number bytes used as control packets to the total number of bytes used for data packets. Percentage of control overhead is calculated for AT-ST scheme and PS2_MAC for all the five Scenarios. For convenience, we show the performance results of observed while adjusting HP in Figure 6. The addition of separate AT and ST packets increases the overall control packet overhead by 9.5%. Result shows the performance of HP in AT-ST Scheme and PS2-MAC. It clearly shows that the control packet overhead of AT-ST Scheme is high especially when the number of HP nodes is more in the network. This is because, there would be more chances of priority reversal and hence number of ST packets transmitted would be more. Further it is observed that there is lot of variation in the control packet overhead caused by AT-ST because whenever there is a priority reversal ST is used. But it is almost the same in PS2-MAC because we do not use any extra control packet and the smaller variation is because of the unused ACK packets when the transmission is suspended. Thus, the result shows that the control overhead is significantly reduced in the proposed approach.

## V CONCLUSION

In this paper, we propose PS2-MAC model which provides prioritization and differentiation based on user profiles. It provides prioritization at three levels. First, it assigns user profile based priority such as HP, MP and LP and stamps the packets accordingly and maintains three queues. Secondly, weights are assigned to three user profiles according to the required proportional differentiation. Apart from weights, the percentage of packets waiting in their respective queues are considered we use rationed dequeuing algorithm to normalize the access ratios and dequeue the number of packets accordingly so that, starvation is avoided for Low priority nodes. Finally prioritization is achieved at the MAC layer through proportional share scheduling which includes, variable inter frame space, proportionate prioritized backoff timers. Contention for channel access is achieved through enhanced RTS/CTS control packets. Comparison of control overhead is made with the AT-ST scheme. Simulation results show that 9.5% control overhead is reduced in our proposed scheme compared to AT-ST scheme. Results also justify differentiated services have been achieved for the different profiles of users thus improving fairness and reducing starvation. Though our model supports dynamicity based on queue length, it focuses on static user profile based weights. As a future work, we plan to enhance this model by introducing dynamic weights.

**Authors**

**J. Hannah Monisha,** is presently heading the Department of Computer Science in Indira Gandhi College of Arts and Science (Government of Puducherry), Puducherry and has over 16 years of professional experience. She has completed M.Sc. Degree in Computer Science, M.S Degree in Information Technology, M. Phil. Degree in Computer Science and currently doing Ph.D in the field of Mobile Ad Hoc Networks. She is a member of the Computer Science curriculum design board of Pondicherry University, India. She is a member of the ACM, CSTA. She is a voracious writer and has published many articles in national and international journals. Her area of interest includes, Parallel Algorithms, Mobile Computing and Social Networks.

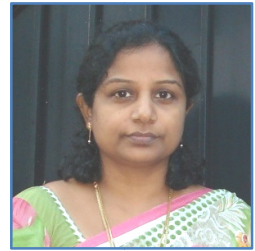

**Dr. V. Rhymend Uthariaraj** is a Professor & Director of Ramanujan Computing Centre, Anna University, Chennai. He holds an additional post of Secretary for Tamil Nadu Engineering Admissions and Co-ordinator and AICTE-MCA QIP Programme at Anna University, Chennai. He holds a Master of Engineering in Computer Science and Engineering and Ph.D in Computer Science and Engineering from Anna University, Chennai. His area of research includes Network Security, Pervasive Computing, Distributed Computing, Operations Research, and Computer Algorithms. He has 27 years of research experience and had guided many scholars. He is a member of Indian Society for Technical Education.

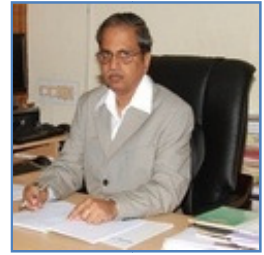